\documentclass{article}

\usepackage{hyperref}
\hypersetup{colorlinks=true, linkcolor=black, filecolor=magenta, urlcolor=black, pdftitle={Overleaf Example}, pdfpagemode=FullScreen}
\usepackage{graphicx}
\graphicspath{ {./img/} }
\usepackage{outlines}
\usepackage{algorithm2e}
\usepackage{enumitem}
\usepackage{float} 
\usepackage{alltt}
\usepackage{url}
\usepackage{xspace}

\RestyleAlgo{ruled}

\SetKwComment{Comment}{/* }{ */}

\usepackage{cite}

\newcommand{\ie}{i.e.,\xspace}
\newcommand{\cd}[1]{\texttt{#1}}
\newcommand{\remove}[1]{}

\headheight     = \baselineskip
\headsep        = \baselineskip
\textheight     = \paperheight
\textwidth      = \paperwidth
\linewidth      = \textwidth

\begin{document}


\title{Code Documentation and Analysis to Secure Software Development}

\author{
Paul Attie, Anas Obeidat, Nathaniel Oh, Ian Yelle\\
School of Computer and Cyber Sciences\\
Augusta University\\
Augusta, Georgia 30912
}

\maketitle

\begin{abstract}

We present the Code Documentation and Analysis Tool (CoDAT).  CoDAT is a tool designed to maintain consistency between the various levels of code documentation, e.g. if a line in a code sketch is changed, the comment that documents the corresponding code is also changed. That is, comments are linked and updated so as to remain internally consistent and also consistent with the code.
By flagging "out of date" comments, CoDAT alerts the developer to maintain up-to-date documentation.

We use a large language model to check the semantic consistency between a fragment of code and the comments that describe it. Thus we also flag semantic inconsistency as well as out of date comments. This helps programers write code that correctly implements a code sketch, and so provides machine
support for a step-wise refinement approach, starting with a code
sketch and proceeding down to code through one or more refinement iterations.

CoDAT is implemented in the Intellij IDEA IDE where we use the Code Insight daemon package alongside a custom regular expression algorithm to mark tagged comments whose corresponding code blocks have changed. CoDAT's backend is structurally decentralized to allow a distributed ledger framework for code consistency and architectural compilation tracking.
\end{abstract}                      

\section{Introduction}

Documenting source code both inline and through external means provides essential insight and continuity for any developer regardless of time spent on the code base itself. While good comments doesn't fix bad code, it's essential to both enhance the readability and interpretability for an application\cite{alibabacloudCodeComments}. It is important to not only comment code, but to comment code well. 

The goal of commented code should be allow a user to implement a program or feature successfully through simply reading the inline comments or external documentation \cite{utahProgrammingCommenting}. Thus, it is important to comment and document a program's development in real-time rather than attempting to retroactively narrative the code's purpose and functionality. Oftentimes, good code documentation consists of the following components\cite{stackoverflowBestPractices}: 

\begin{itemize}
    \item Can plainly describe the code without needing to include the code itself.
    \item Does not increase the complexity of the program, but rather simplifies it.
    \item Includes any future changes or bug fixes.
    \item Provides references to further reading or additional documentation when needed.
\end{itemize}

While external documentation serves as a important central resource for developers and users alike, inline code commenting is essential for development continuity and debugging\cite{codinghorrorCodeTells}. 
Thus, there remains a need to enforce documentation linking for both developers and reverse engineers. In this paper, we will cover CoDAT from both a developer's and from a reverse engineer's point of view.

\subsection{Code Reviews}

Code reviews are an important milestone in the life cycle of any major software project \cite{gitlabWhatCode}. Traditional code reviews take on a more rigid and formal approach, focusing on depth and coverage rather than speed and scalability \cite{Fagan76,Fagan86}.
Historically, code reviews have been effective in finding software bugs \cite{BacchelliBird13}, \cite{BasiliSelby87}. 

However, as code size has increased, the effectiveness of code reviews has correspondingly decreased.
Czerwonka et. al. show that the traditional approach of formally reviewing code and providing feedback is negatively correlated with the size of the code review \cite{CzerwonkaEtAl15}. Therefore, the implementation of small-scale just-in-time code reviews have shown greater probably of success than their traditional large-scale counterparts \cite{codeReviewForAndBy}.

While there are many variables that may impact the quality of a code review, the biggest factor is the thoroughness of the feedback, the reviewer's familiarity of the code, and the quality of the code itself \cite{Kononenko16}. This in and of itself is important to keep in mind as this shares parallels with what makes a good inline code comment \cite{stackoverflowBestPractices}. As we explore the crossroads between change management and comment tracking, it's important to always keep in mind that the reviewer defines the quality and the review is just a means of enforcing it. 

Another important aspect to discuss about code reviews is the influence code coverage has on the code review's quality \cite{McIntosh14}. Emphasizing code coverage is an important link that can be easily maintained through just-in-time small but frequent code reviews. With this new focus on the small-scale approach to code reviews, up-to-date and accurate inline code comments are now more important than ever. For example, code review comments and updates may be incorporated into inline comments to describe a particular change or implementation\cite{NikithaEtAl22}. Yet, this effort may fall short if users are not routinely made aware of inline comments that need to be updated. 

Manual review of code by professional programmers tend to perform better than automated or procedural approaches\cite{BasiliSelby87}. However, programs continually grow, their code bases becoming increasingly complex and large. As such, manual code reviews are becoming harder to scale and thus more nuanced and novel approaches are needed. 

\subsection{Code Quality}

A program's quality may very greatly depending on a programmer's technical background and ability to perform\cite{Dieste2017}. Code quality can be measured by the number of bugs found within a program\cite{designSpaceOfBugFixes}. The greater the number of bugs present, the less the program will perform as intended and will more likely be susceptible to external attacks. Thus, it is important to enforce and quantify a program's quality to prevent these attacks and reduce the number of bugs present. 

One of the metrics historically used to track quality of code is through the quality of its comments\cite{9610657}. While not measured as a direct one-to-one correlation, existing research assesses code quality through regression testing\cite{10.1145/367008.367020} and code smells\cite{10.1007/s11334-013-0205-z}. While approaches may vary across domain implementation, the program attributes themselves need to adhere to adequate specification in order to be considered of high quality\cite{LEHMAN198419}.

\subsection{Integrated Development Environments, Plugins, and Change Tracking}

An Integrated Development Environment, or IDE\cite{amazonWhatIDE}, is traditionally defined as a software application that assists developers through various code editing, compiling, and checking techniques. However, oftentimes a particular IDE may not have specific functionality a developer wants when it comes to a niche application or problem. 

This is often solved through the implementation of 3rd party helper applications called plugins\cite{lenovoPluginWhat}. In our case, we created a plugin for the Java and Kotlin IDE, Intellij IDEA\cite{jetbrainsIntelliJIDEA}. 
Our plugin integrates a hyperlink tree structure with state-aware change tracking functionality. 

Traditionally, change tracking is handled by programs such as Subversion\cite{apacheApacheSubversion} or Git\cite{gitscm}. Our approach takes the incremental change tracking and updates the user in real-time when a particular change may invalidate the affiliated code sketch or comment.

\section{CoDAT: Our Approach to Code Documentation}

We advocate code documentation at many levels of abstraction, so that documentation has a hierarchical structure. A key goal is to maintain consistency between abstraction levels. CoDAT provides automated support for this by tracking changes and alerting the developer to related documentation at higher/lower levels. Also, an LLM can be invoked to check consistency between documentation and code, and also between documentation at ``adjacent'' abstraction levels.

The documentation structure that we propose is as follows:
\begin{itemize}
\item Top tier of documentation: functional specification of modules,
  i.e., what is a class or method required to do?
    \begin{itemize}
    \item Classes: Purpose of the class, the major data structures.
    \item Methods: a functional specification, given by two clauses:
        \begin{enumerate}
        \item Requires clause: constraints on inputs
        \item Effects clause: states what the module does
        \end{enumerate}
    \end{itemize}
\item Second tier of documentation: \emph{code sketching}, how does code work, at a ``high'' level.
A code sketch expresses the key algorithm underlying the code without getting bogged down in coding details.
\item Third tier: in-line code comments, more detailed description of the code, details of data structures and algorithms.
\item Further levels add detail, until a level is reached that is straightforward to translate into working code.
\item Number of levels needed depends on the task complexity.
\end{itemize}

Code sketching provides a form of a Documentation for both the code itself and the next level of code abstraction. Code sketching can be used in code reviews to better help developers understand the code. CoDAT implements the above vision of code documentation by providing the following functionality:

\begin{itemize}
\item Automated management of documentation
\item Change flagging:  change in code/documentation is linked to related documentation that may need to be changed, and user is alerted
\item Consistency checking: LLM is used to check that comments and corresponding code match w.r.t. described functionality. Serves as a form of soft verification.
\item Views: documentation serves as code blueprint at various levels of abstraction
Implemented in the Intellij IDEA Development Environment
\end{itemize}

Our proposed documentation structure helps maintain a ``mental image'' of how the code works, and 
therefore helps with the cognitive workload of:

\begin{itemize}
\item Tracing the code
\item Figuring out how the code works and 
\item How different data structures and methods relate
\item Makes debugging faster, easier, better
\end{itemize}

CoDAT provides a way to formalize and partially automate small-scale code reviews that support a larger base of code. In particular, we see to leverage CoDAT to automate consistency and regression checking via effective code documentation. While CoDAT is functionally independent from machine learning and artificial intelligence, it does have the ability to integrate with a third-party LLM. 

With an LLM, CoDAT can significantly reduce the manual labor needed to perform consistency checks and regression analysis. The pinpoints for these checks are often ad-hoc and in the spur of the moment. A particular feature or file may change a countless number of times prior to a formal code review. Thus, having a more informal and dynamic approach is preferred. 

Given this dynamic approach, we seek to model CoDAT's behavior off of existing source control systems such as Git or Subversion. Our focus with CoDAT is to explore the junction between code documentation and stateful change tracking. Oftentimes developers may change the functionality of a program without updating the inline comments or code sketches. While wikis may serve as a substitute for some developers, we believe that inline code commenting is an essential practice in any code base, especially large projects. 

\begin{figure}[!htp]
\centering
\includegraphics[width=\linewidth]{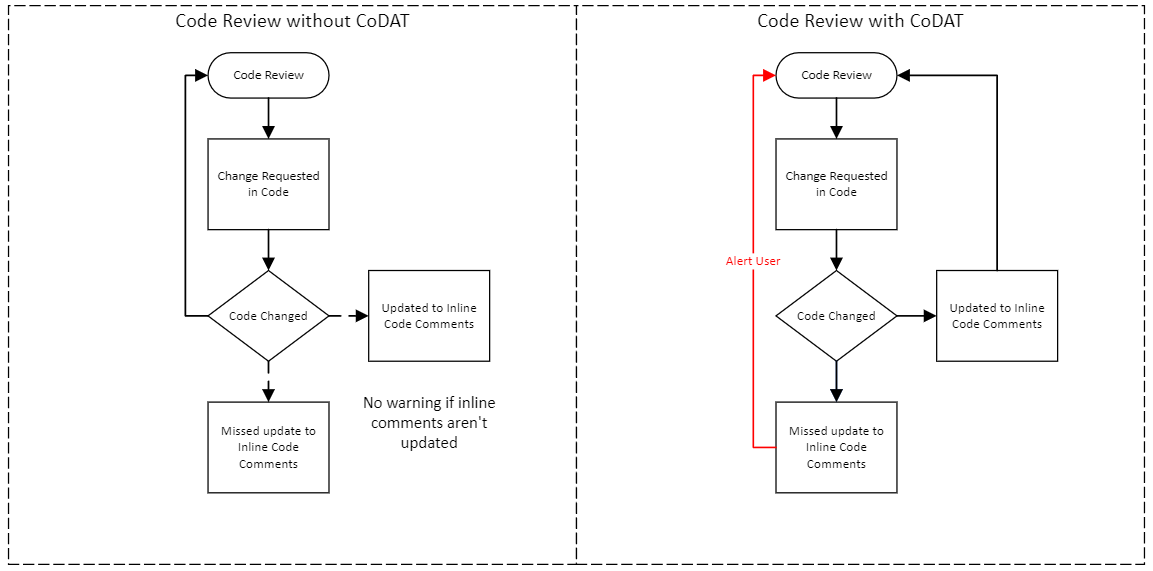}
\caption{Code Reviews with CoDAT}
\label{fig1}
\end{figure}

As shown in Figure \ref{fig1}, CoDAT provides developers with the ability to be notified in real-time when inline code comments or sketches needs updating. While CoDAT does not block the user from not updating the inline comment or sketch, CoDAT will persistently alert the user that a code change occurred for a related code sketch or comment.

\section{CoDAT Implementation and Future Development}


A key aim of CoDAT is to bridge the semantic gap between what the code appears to do and its actual
behavior, reducing the chances of bugs and enhancing documentation quality.
CoDAT is implemented as a plugin which integrates seamlessly into IntelliJ IDEA, managing hierarchical documentation structures and linking comments to related code.
The main functionality provided is:
\begin{itemize}
\item Parse and identify comments in the source code.
\item Manage and store comments using a hierarchical data structure.
\item Provide functionalities like highlighting, navigating, and updating comments.
\item Interface with the IDE's text editor for displaying and interacting with comments.
\end{itemize}

\subsection{Overview of CoDAT Architecture and Main Data Structures}

\begin{figure}[!htp]
\centering
\includegraphics[width=\linewidth]{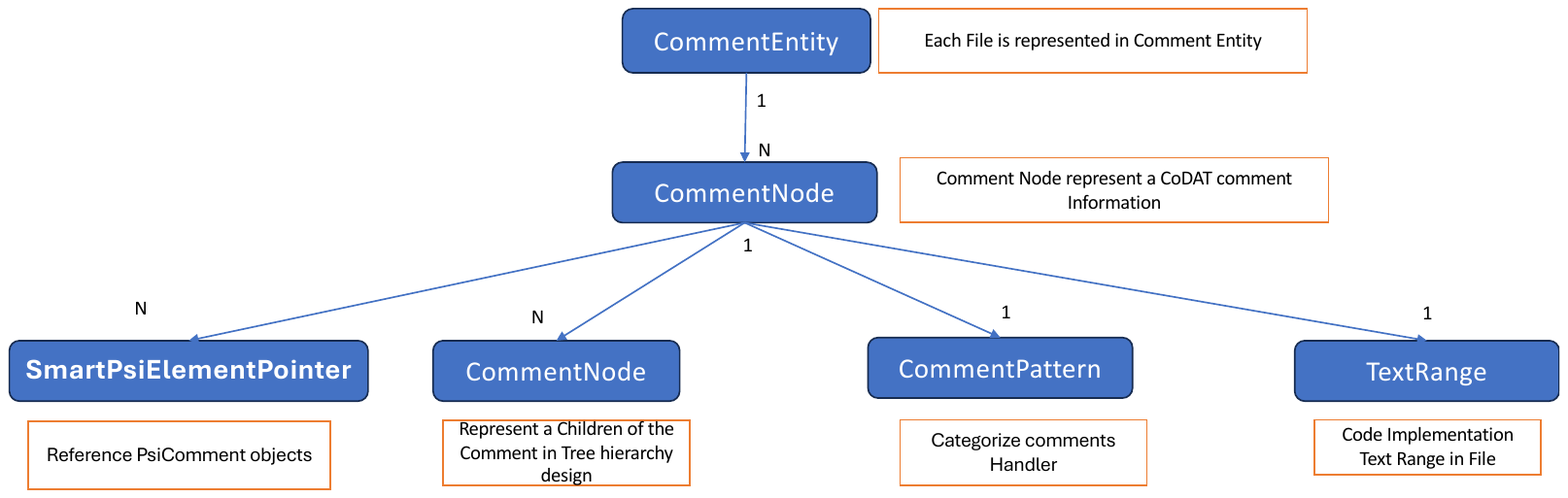}
\caption{Main data structures}
\label{fig:dataStructs}
\end{figure}

The CoDAT plugin’s architecture is modular, leveraging IntelliJ’s Program Structure Interface (PSI) for seamless integration. 
Key components are structured into distinct layers:

\begin{enumerate}
\item IDE Environment Layer: Provides fundamental services like text editing, project management, and version control. The CoDAT plugin operates within this environment.
\item Plugin System Layer: Comprises various modules that handle parsing, management, and user interaction:
   \begin{itemize}
   \item Comment Parsing Module: Analyzes source files, extracts comments, and categorizes them hierarchically.
   \item Comment Management Service: Handles the lifecycle of comments across files, linking them to associated code blocks and ensuring consistency.
   \item UI/UX Component: Offers an intuitive user interface for visualizing hierarchical documentation and navigating code.
   \end{itemize}

\item Core Data Structure Layer: Contains the main data structures representing hierarchical
  comments.Figure~\ref{fig:dataStructs} gives a data model for the current
  CoDAT implementation.
  \begin{itemize}
    \item \cd{CommentEntity}: A container mapping files to their
      corresponding comment trees.
      Acts as the root or container for managing multiple comment
      threads or trees. It can hold multiple \cd{CommentNode} instances,
      each corresponding to different files or sections within files.
      
    \item \cd{CommentNode}: Represents individual comments or groups of comments. 
          Forms a hierarchical tree structure, since a CommentNode can contain several other
          CommentNodes, which are its children in the comment tree.
          This is useful for nested (\ie multi-level) comments.

   \item \cd{SmartPsiElementPointer}: Used within \cd{CommentNode} to safely reference \cd{PsiComment} objects even as the underlying code changes.

   \item \cd{CommentPatterns}: Used to categorize or match comments within the nodes
based on predefined patterns.

   \item \cd{TextRange}: object could be designed to encompass the start and end points of a text block, making it highly relevant in applications involving syntax highlighting, code analysis, and document editing features. 
          
   \end{itemize}
 \end{enumerate}

\remove{
\begin{alltt}
public class CommentEntity {
private Map<String, Map<String, CommentNode>> fileComments;
// Initializes an empty structure for managing multiple files.
public CommentEntity() {
this.fileComments = new HashMap<>();
}

// Adds a new CommentNode to the map for a specific file.
public void addComment(String fileName, String label, CommentNode node) {
fileComments.computeIfAbsent(fileName, k -> new HashMap<>()).put(label, node);
}

// Retrieves all comments for a particular file.
public Map<String, CommentNode> getCommentsForFile(String fileName) {
return fileComments.getOrDefault(fileName, new HashMap<>());
}}
\end{alltt}      
}

\remove{
\begin{alltt}
public class CommentNode {
private String label;
private List<CommentNode> children;
private List<SmartPsiElementPointer<PsiComment>> psiComments;

// Initializes a comment node with the given label and empty children.
public CommentNode(String label) {
this.label is label;
this.children = new ArrayList<>();
this.psiComments = new ArrayList<>();
}

// Adds a child comment node.
public void addChild(CommentNode child) {
children.add(child);
}

// Links a PSI comment to this node.
public void addPsiComment(SmartPsiElementPointer<PsiComment> psiComment) {
psiComments.add(psiComment);
}

// Retrieves all children of this node.
public List<CommentNode> getChildren() {
return children;
}

// Gets the PSI comments linked to this node.
public List<SmartPsiElementPointer<PsiComment>> getPsiComments() {
return psiComments;
}}
\end{alltt}
}

\subsection{Lessons learned}

\begin{enumerate}

\item Hierarchical Documentation is Crucial

Organizing comments hierarchically helps improve navigation and makes code reviews more consistent.
The CoDAT data structure implemented in \cd{CommentEntity} and \cd{CommentNode} organizes comments in a logical tree-like 
format, which simplifies the process of tracking changes and managing code blocks.

\item Change Flagging Adds Value

Automatically flagging changes in code and comments ensures the documentation remains consistent with the actual implementation.
However, fine-tuning this feature to balance sensitivity is crucial. Too many alerts can overwhelm developers, while too few may overlook important changes.

\item Deep IDE Integration is Challenging
  
IntelliJ’s PSI (Program Structure Interface) API is powerful but complex. Integration requires a deep understanding of how IntelliJ represents program elements.
While implementing the \cd{CommentLinkMarkerProvider} and \cd{CommentEditListener}, we found that maintaining accurate references to PSI elements is challenging because code constantly evolves.

\item UI/UX Design Requires Early Prototyping

The UI/UX components that manage hierarchical comments need thorough user testing.
During development, implementing the graphical interface for comment visualization required frequent adjustments. Early prototyping would have provided valuable user feedback to refine the design and improve the interaction flow.

\item Modular Code Design Facilitates Extension

Designing the plugin as modular components makes it easier to extend.
By separating the comment parsing, management, and UI modules, the development team can work on different aspects simultaneously and future extensions can be implemented with minimal disruptions.

\item Comprehensive Testing is Essential

Changes in hierarchical comments can have ripple effects on the documentation structure.
Implementing unit and integration tests across \cd{CommentHandler}, \cd{CommentEntity}, and \cd{CommentNode} is necessary to ensure accurate parsing, consistent linking, and proper navigation.

\item Team Collaboration Enhances Documentation Standards

CoDAT enables developers to collaborate on documentation across multiple abstraction layers.
However, creating a unified documentation standard for the team ensures consistency in coding practices and helps identify anomalies effectively.

\end{enumerate}

\subsection{Future Work}

\begin{enumerate}

\item Advanced Signature Detection

  Implement a robust change detection system that evaluates code changes and their impact across the entire application.
  Consider integrating static analysis techniques to identify affected areas in real-time and suggest necessary documentation updates.

\item NLP-Based Comment Summarization

Leverage natural language processing (NLP) models to analyze comments and generate accurate summaries.
Develop a recommendation system that suggests changes to comments based on new code patterns.

\item Enhanced Visualization

Develop richer visualizations and interactive components to explore the hierarchical comment structure.
Implement ``heat maps'' to identify areas of code that frequently change and ensure their
documentation aligns with recent modifications. Introduce a timeline view to visualize the evolution of comments and code over time.

\item Version Control System Integration

Deeply integrate with source control systems (Git, Subversion) to track and compare changes in code and comments automatically.
Build a dashboard that links changes to their corresponding version control history, allowing developers to audit the development process.

\item Machine Learning-Based Anomaly Detection

Develop machine learning models to detect inconsistencies or anomalies in the documentation and
code, flagging possible areas needing attention. Train models to predict common documentation pitfalls, suggesting areas that may require further reviews.

\item Code Review Automation

Automate the code review process further by developing rule-based and AI-driven tools to review documentation and ensure comments align with project guidelines.
Implement suggestion features that highlight areas where comments can be improved or expanded.

\item Customized Project Templates

Create project templates with pre-defined documentation standards for different development teams.
Allow teams to customize templates based on their unique workflows and coding guidelines.

\end{enumerate}
  

\section{Integrating a Large Language Model into CoDAT}

CoDAT has an LLM backend to perform consistency analysis for code documentation.
We use the Claude Haiku LLM \url{anthropic.com}.
We use a Python script that takes an HTTP request through localhost to then send to the LLM as
before and return the response to the HTTP request using the Python Flask and Requests libraries
 (claude\_chat\_v1.py).
This requires an API key to use and communicates to
Anthropic’s Claude AI. It uses the Python Flask and Requests libraries and is documented thoroughly in the program itself.
We use a simple Python script that takes in an input and sends it on localhost as a POST HTTP request
(test\_send.py). This also uses the Python Requests library.

\subsection{Future work on LLM integration into CoDAT}

We will attempt different ways of mitigating the LLM’s tendency for false
results, including but not limited to a) using multiple unique LLMs for redundancy and lessened chance of a fluke, b)
using formal methods and more traditional logic to double check the LLMs output or to send formal methods proofs to the
LLM rather than code and comments, or c) develop or find a more fault-tolerant LLM or a way of wording inputs that
results in less false results. In addition, the use of LLMs for proving formal methods or the development of a formal
methods trained LLM is not something I’ve heard about or found through my research so that might be a field worth
pursuing. Hopefully this helps bring more fruitful research in the years to come.


\section{CoDAT Interface}

\begin{figure}[htp]
\centering
\includegraphics[width=\linewidth]{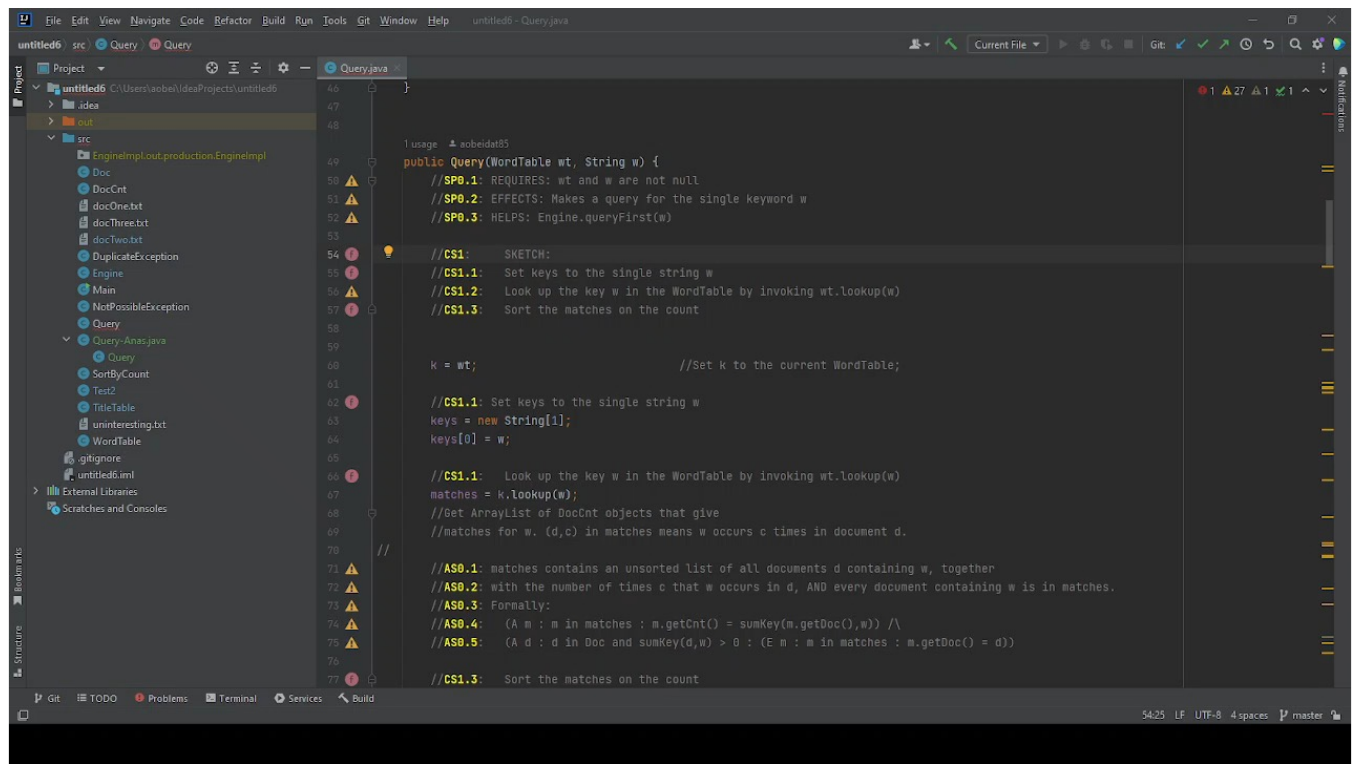}
\caption{CoDAT Interface}
\label{fig:interface}
\end{figure}

Figure~\ref{fig:interface} shows the CoDAT interface.
The main features of the CoDAT interface are as follows:

\subsection{Click Navigation}

\paragraph{Direct Navigation to Code} Click on any node within the CoDAT tool window to seamlessly navigate to the corresponding code block in the IntelliJ editor. This interaction leverages IntelliJ’s powerful navigation capabilities to ensure that developers are taken directly to the relevant section of the code.

\paragraph{Automatic Highlighting} Upon navigation, the associated comment and the entire code block are automatically highlighted. This visual cue helps developers quickly understand the context and purpose of the code segment, enhancing their ability to review changes effectively.

\subsection{Gutter Icons}

\paragraph{Visual Indicators} Gutter icons, which appear as small symbols in the editor’s left margin, serve as visual indicators of where comments are located within the code file.

\paragraph{Interactive Icons} These icons are interactive, allowing developers to click on them to navigate directly to the corresponding node in the CoDAT tool window. This feature is particularly useful for visual learners and enhances the ability to trace relationships and dependencies across the codebase.

\paragraph{Streamlined Workflow} The integration of gutter icons simplifies the process of navigating complex code files, making it easier to manage large projects with extensive documentation.

\subsection{Highlighted Comment Blocks}

\paragraph{Enhanced Code Visibility} Clicking on a node or gutter icon not only navigates to the associated comment but also highlights all lines of code linked to that comment. This comprehensive highlighting approach ensures that developers can see at a glance all parts of the code that are discussed or described by the comment.

\paragraph{Impact Analysis} Highlighted comment blocks help developers quickly assess the impact of specific sections of code. This is essential for understanding how changes in one part of the code might affect other parts, facilitating more effective debugging and code modification.

\paragraph{Documentation Consistency Checks} By highlighting the related code segments, developers are better equipped to verify that the documentation remains accurate and up-to-date with the code’s current functionality. This feature supports ongoing maintenance and helps prevent code decay.

\subsection{Annotating Code}

\paragraph{Adding Annotations}
For inline comments, insert inline comments using // syntax and follow a pattern like //CS1.1: to categorize it. Example:

\begin{verbatim}
//CS1.1: Initialize the database connection
Database db = new Database(config);
\end{verbatim}

For block comments, use block comments (/* ... */) for more comprehensive descriptions. Example:

\begin{verbatim}
/*CS2.1: Module Configuration
This section sets up the necessary parameters for
the module to function as intended.
*/
\end{verbatim}

We use CS1, CS2 etc. to label comments within a code sketch, and AS1, AS2, etc to label Hoare-logic assertions.

\paragraph{Managing Annotations}
The CoDAT tool window organizes and displays comments hierarchically, allowing you to navigate and manage them efficiently.
Right-click on a node in the tool window to edit or remove comments.

\section{Example: Document Search Engine}

We present an example application of CoDAT. The program to be documented and analyzed is a
\emph{document search engine}. There is a given collection of \emph{documents}, which can be added to. A \emph{query} consists of a sequence $w_1,\ldots,w_k$ of keywords. A document matches if it contains all the keywords. Matches are ranked by the
\emph{total occurrence count}: the sum of the occurrences of all the keywords.
The required functionality is as follows:
\begin{itemize}
\item Start a new query, by giving an initial keyword $w_1$
\item Add a keyword $w_k$ to an existing query $w_1,\ldots,w_{k-1}$
\item Add a new document to the current collection. If a query is in progress, update the query with the new document, if the new document is a match.
\end{itemize}

We give the search enginre code in
Appendix~\ref{app:docSearchCode}. The main classes are as follows:
\begin{itemize}
\item \cd{Engine}: top level class that provides the client interface.
\item \cd{Query}: creates and updates a \cd{query} object, which
  stores the list of currently matching documents, sorted by total
  occurrence count.
\item \cd{DocCnt}: a pair consisting of a document and a count (\ie an integer) 
 \item \cd{WordTable}: maintains a mapping from words to lists of
  \cd{DocCnt} objects. $(d,c)$ in the list of $w$ means that $w$
  occurs $c$ times in document $d$.
  \item \cd{Doc}: represents a document; stores the title and the body
    of the document, both as strings.
\item \cd{TitleTable}: maintains a mapping from titles to their
  corresponding documents.
\end{itemize}


\section{Conclusion}

The CoDAT project provides a comprehensive tool for enhancing code reviews and ensuring consistent documentation. The IntelliJ plugin efficiently organizes hierarchical comments, linking them to related code and managing their lifecycle. Future work will involve improving signature detection and LLM-based analysis while integrating more deeply with source control systems to provide seamless documentation tracking.

A key advantage of our approach is that it is not subject to known limitations of AI such as hallucinations, since we do not use an LLM to generate code, but only to check the consistency of code versus its documentation (code sketches and inline comments/assertions).
The LLM thus acts as an assistant to the programmer, and the code summaries that it generates can be helpful in checking the code.
We expect that incorrect/unexpected responses by the LLM will lead to carefull review and debugging by the programmer.

\clearpage 
\appendix


\section{CoDAT User Manual}

\subsection{Installation}

•	Prerequisites: Make sure you have IntelliJ IDEA (Community or Ultimate Edition) installed.
•	Download and Install the Plugin:

Manual Installation:
\begin{enumerate}
\item Download the CoDAT plugin .zip file or .jar file.
\item In IntelliJ IDEA, go to File $>$ Settings $>$ Plugins.
\item Click ``Install Plugin from Disk'' and choose the downloaded file.
\item Restart IntelliJ IDEA to complete the installation.
\end{enumerate}

\subsection{Basic Usage:  Navigating and Highlighting Comments}

\subsubsection{Click Navigation}
\begin{itemize}
\item Direct Navigation to Code: Click on any node within the CoDAT tool window to seamlessly navigate to the corresponding code block in the IntelliJ editor. This interaction leverages IntelliJ’s powerful navigation capabilities to ensure that developers are taken directly to the relevant section of the code.
\item Automatic Highlighting: Upon navigation, the associated comment and the entire code block are automatically highlighted. This visual cue helps developers quickly understand the context and purpose of the code segment, enhancing their ability to review changes effectively.
\end{itemize}

\subsubsection{Gutter Icons}
\begin{itemize}
\item Visual Indicators: Gutter icons, which appear as small symbols in the editor’s left margin, serve as visual indicators of where comments are located within the code file.
\item Interactive Icons: These icons are interactive, allowing developers to click on them to navigate directly to the corresponding node in the CoDAT tool window. This feature is particularly useful for visual learners and enhances the ability to trace relationships and dependencies across the codebase.
\item Streamlined Workflow: The integration of gutter icons simplifies the process of navigating complex code files, making it easier to manage large projects with extensive documentation.
\end{itemize}

\subsubsection{Highlighted Comment Blocks}
\begin{itemize}
\item Enhanced Code Visibility: Clicking on a node or gutter icon not only navigates to the associated comment but also highlights all lines of code linked to that comment. This comprehensive highlighting approach ensures that developers can see at a glance all parts of the code that are discussed or described by the comment.
\item Impact Analysis: Highlighted comment blocks help developers quickly assess the impact of specific sections of code. This is essential for understanding how changes in one part of the code might affect other parts, facilitating more effective debugging and code modification.
\item Documentation Consistency Checks: By highlighting the related code segments, developers are better equipped to verify that the documentation remains accurate and up-to-date with the code’s current functionality. This feature supports ongoing maintenance and helps prevent code decay.
\end{itemize}

\subsection{Basic Usage:  Annotating Code}

\subsubsection{Adding Annotations}

\begin{itemize}

\item Inline Comments: Insert inline comments using \cd{//} syntax and follow a pattern like \cd{//CS1.1:} to categorize it. Example:

  \cd{//CS1.1: Initialize the database connection}\\
  \cd{Database db = new Database(config);}
  
\item Block Comments: Use block comments \cd{(/* ... */)} for more comprehensive descriptions. Example:
\begin{alltt}
  /*CS2.1: Module Configuration
This section sets up the necessary parameters for
the module to function as intended.
*/
\end{alltt}

\end{itemize}

\subsubsection{Managing Annotations}

\begin{itemize}
\item The CoDAT tool window organizes and displays comments hierarchically, allowing you to navigate and manage them efficiently.
\item Right-click on a node in the tool window to edit or remove comments.
\end{itemize}

\subsubsection{Advanced Configuration}

Custom Comment Patterns: Create custom patterns to categorize comments according to your team’s requirements. For instance, patterns like
\cd{//SP:} or \cd{//TODO:} can be added to CoDAT’s settings.


\section{Source Code for the Document Search Engine}
\label{app:docSearchCode}

\subsection{Source Code for the Engine Class}

\begin{verbatim}
  
public class Engine {
    //OVERVIEW: An engine has a state as follows:
    //1. A collection of documents to be searched, which is an implicit input for all methods.
    //   Documents can be added to the collection.
    //   Each document consists of a title and a body. Each of these is a sequence of words.
    //   A word is an alphabetic string.
    //2. A sequence of keywords, each resulting from a sequence of submitted queries.
    //3. A list of documents matching all of the keywords (ie. every keyword occurs
    // at least once in every match) and
    //   sorted by total number of occurrences of all keywords.
    //4. A private file ("uninteresting.txt") in the same directory contains the
    // uninteresting words.
    //
    //   Methods throw the NotPossibleException when there is a problem.
    //   For efficient search, documents are maintained in summarized form.


    private WordTable wt; //Summarizes the documents for fast search
    private TitleTable tt; //Enables lookup of documents by title
    private Query q;
    //  private String urls; //WILL IGNORE URLS IN PROTOTYPE SINCE WE READ DOCS FROM FILES


    //CONSTRUCTORS

    public Engine()   throws NotPossibleException, FileNotFoundException {
        //EFFECTS: If the uninteresting words cannot be read from the private file
        //throws NotPossibleException, else creates NK and initializes the
        //application state appropriately.

        //IMPL SKETCH:
        //wt := WordTable()
        //tt := TitleTable()
        //q := null
        //urls is initially empty.

        wt = new WordTable();
        tt = new TitleTable();
        q = null;

    }



    //METHODS

    public Query queryFirst(String w) throws NotPossibleException {
        //EFFECTS: If w = null or not WORD(w) or w in NK throws
        //NotPossibleException else
        //sets Key = {w}, performs the query, and returns the result.

        //IMPL SKETCH:
        //if precondition is false then raise exception else q := new Query(wt, w)

        if (!wt.isInteresting(w)) throw new NotPossibleException("Engine.queryFirst");
        else {
            q = new Query(wt, w);
            return q;
        }

    }


    public Query queryMore(String  w) throws  NotPossibleException {
        //EFFECTS: If  w = null or not WORD(w) or w in NK or
        //   Key = emptyset or w in Key throws NotPossibleException,
        //   else adds w to Key and returns the query result.
        //IMPL SKETCH:
        //if precondition is false raise exception else
        //     add w to keys[]
        //     q := q.addKey(w)

        if (!wt.isInteresting(w) ) throw new NotPossibleException("Engine.queryMore");  

        String[] keys = q.keys();

        if (keys.length == 0) throw new NotPossibleException("Engine.queryMore"); 
        //Key = emptyset
        //{WORD(w) /\ w notin NK /\ keys != emptyset}

        for(int i=0; i < keys.length; i++)
            if (w==keys[i]) throw new NotPossibleException("Engine.queryMore");
        //{WORD(w) /\ w notin NK /\ keys != emptyset /\ w notin keys}

        q.addKey(w);
        return q;
    }




    public Query addDocFromFile(String f) throws NotPossibleException, FileNotFoundException,
    DuplicateException {

        //EFFECTS: If f is not a name for a local file (in the same directory) that:
        //   (1) can be opened for reading and
        //   (2) whose contents can be interpreted as a document
        //  then throws NotPossibleException,
        //  else adds the new documents to Doc.
        //  If no query was in progress returns the empty query,
        //  else returns the query result that includes any matching new documents.

        //IMPL SKETCH:
        //open file f for reading
        //read in and construct document d
        //tt.addDoc(d), add d to the title table
        //Hashtable h := wt.addDoc(d) is a hashtable for document d only
        //if q != null then q := q.addDoc(d,h) else q := new Query()
        //return q

        java.io.File docFile = new java.io.File(f);
        Scanner input = new Scanner(docFile);

        String s =  input.useDelimiter("\\A").next();    //read all of Scanner input, since
        // \A is boundary marker for beginning of input

        Doc d = new Doc(s);                //Construct document d corresponding to file f
        tt.addDoc(d);                      //Add d to TitleTable.
        Map h = wt.addDoc(d);              //Add d to WordTable and get hashtable h for d only

        if (q != null) {
            q.addDoc(d,h);      //Update ongoing query q with d
            return q;
        }
        else {
            return(new Query());        //If no ongoing query, return empty query.
        }
    }
}
\end{verbatim}

\subsection{Source Code for the Query Class}

\begin{verbatim}

public class Query {

    //CS1 OVERVIEW: Maintains the keywords w_0,...,w_{k-1} of a query and the list of documents
    //CS2  that match those keywords. The list is ordered by total occurrence count, i.e,
    //CS3 sum of occurrences of w_0,...,w_{k-1} in each document. Also each of w_0,...,w_{k-1}
    // must occur at least once in each matching document.
    // size returns the number of matches.
    // Documents can be accessed using indexes between 0 and size-1 inclusive.
    // Documents are ordered by the number of matches they contain,
    // with document 0 containing the most matches.

    private WordTable k;
    private ArrayList matches; //ArrayList of DocCnt objects. Replaces Vector used in the design.
    private String[] keys;  //The keywords used in the current query

    //Constructors

    public Query() throws NotPossibleException, FileNotFoundException {
        //EFFECTS: Returns the empty query.

        k = new WordTable();                 //WordTable containing only the uninteresting words.
        matches = new ArrayList<DocCnt>();           //empty ArrayList
        keys = new String[0];                //empty Array

    }


    public Query(WordTable wt, String w) {
        //SP1: REQUIRES: wt and w are not null
        //SP2: EFFECTS: Makes a query for the single keyword w
        //SP3: HELPS: Engine.queryFirst(w)

        //SKETCH:
        //CS0:   Set keys to the single string w
        //CS1:   Look up the key w in the WordTable by invoking wt.lookup(w)
        //CS2:   Sort the matches on the count


        k = wt;                              //Set k to the current WordTable;

        //CS0: Set keys to the single string w
        keys = new String[1];
        keys[0] = w;

        //CS1:   Look up the key w in the WordTable by invoking wt.lookup(w)
        matches = k.lookup(w);
        //Get ArrayList of DocCnt objects that give
        //matches for w. (d,c) in matches means w occurs c times in document d.

        //AS0: matches contains an unsorted list of all documents d containing w, together
        //AS1: with the number of times c that w occurs in d, AND every document containing w
        //is in matches.
        //AS2: Formally:
        //AS3:   (A m : m in matches : m.getCnt() = sumKey(m.getDoc(),w)) /\
        //AS4:   (A d : d in Doc and sumKey(d,w) > 0 : (E m : m in matches : m.getDoc() = d))

        //CS2:   Sort the matches on the count
        Collections.sort(matches, new SortByCount());

        //Assert: as above but now matches is sorted on the occurrence count c of w
        //in each matching document d

    }


    // METHODS



    void addKey(String w) throws NotPossibleException {
        //REQUIRES: w is not null
        //MODIFIES: this
        //EFFECTS: If this is empty or w in Key throws NotPossibleException else
        //   modifies this to contain the query for Key union {w},
        //   i.e., w plus the keywords already in the query.
        //HELPS: Engine.queryMore(w)
        //IMPL:
        //CS0:  Throw exception if w is not a valid key for addition to this
        //CS1:  Add w to Key.
        //CS2:  Look up w in the WordTable and store the information about
        //       matches on w in a hash table H
        //CS3:  For each current matching document d:
        //       look up d in the hash table H and
        //       if it is there, store it in an ArrayList A with updated total occurrence count
        //       sort A
        //       set current matches to A


        //CS0:  Throw exception if w is not a valid key for addition to this
        if (matches.size()==0) throw new NotPossibleException();  //this is empty
        for(int i = 0; i < keys.length; i=i+1)
            if (w.equals(keys[i])) throw new NotPossibleException();  //w already in Key

        //Now modify this to contain the query for keys union {w}.

        //CS1:  Add w to keys
        String[] newKeys = new String[keys.length+1];  //new array is one position larger
        for(int i = 0; i < keys.length; i=i+1)         //copy over old keys
            newKeys[i] = keys[i];
        newKeys[keys.length] = w;                      //add w
        keys = newKeys;                                //assign reference to keys array

        //CS2:  Look up w in the WordTable and
        //For each document d in which w occurs c times, 
        //store (d,c) in a hash table H, with d’s title as the lookup key


        //Level 2 impl. sketch
        //let l = (d_0,c_0)...(d_{n-1},c_{n-1}) be the result of looking up w in WordTable
        //declare a hash table H
        //for each element (wd,wc) of l
        //     store (wd,wc) in H using d.title() as the key

        ArrayList l = (ArrayList) k.lookup(w);
        Map H = new HashMap();
        for (int i=0; i < l.size(); i=i+1) {
            DocCnt w_dc = (DocCnt) l.get(i);
            H.put(w_dc.getDoc().getTitle(), w_dc);
        }


        //CS3:  For each current match,
        //    look up the document in the hash table H and
        //    if it is there, store it in an ArrayList
        //  Sort the ArrayList

        //Level 2 impl. sketch
        //declare an arraylist A of DocCnt
        //for each (m_d, m_c) in matches, i.e., each current match
        //    look up m_d.title() in H, which is the hashtable computed above for 
        //         matches on w only
        //    if H(m_d.title()) = (w_d, w_c), then add (m_d, m_c + w_c) to A
        //    else do nothing //H(m_d.title()) = null, so w does not occur in m_d
        //Sort A on the count field

        ArrayList<DocCnt> A = new ArrayList<DocCnt>();
        for (int i=0; i < matches.size(); i=i+1) {
            DocCnt m_dc = (DocCnt) matches.get(i);
            Doc m_d = m_dc.getDoc();
            int m_c = m_dc.getCnt();
            DocCnt H_dc = (DocCnt) H.get(m_d.getTitle());  //H(m_d.title())
            if (H_dc != null) {
                DocCnt New_dc = new DocCnt(m_d, m_c + H_dc.getCnt());
                A.add(New_dc);
            }
        }
        //Sort A on the count field
        Collections.sort(A, new SortByCount());

        //Update matches
        matches = A;

    }





    void addDoc(Doc d, Map h) {
        //REQUIRES: d is not null and h maps strings (the
        //   interesting words in d) to integers (the occurrence
        //   count of the word in d)
        //MODIFIES: this
        //EFFECTS: If each keyword of this is in h,
        //              adds d to the matches of this.
        //HELPS: Engine.addDocs(u), Engine.addDocFromFile(f)
        //Code sketch:
        //CS1:   If the map h for document d contains all the keywords,
        //CS2:        compute the total occurrence count sum for all keywords and
        //CS3:        insert the <d,sum> pair in the ArrayList of matches.
        //CS4:    Otherwise leave matches unchanged.
        //CS5:    In either case, return the resulting query.


        //CS1:   If the map h for document d contains all the keywords,
        boolean b = true;
        for(int i = 0; i < keys.length; i=i+1)
            b = b && h.containsKey(keys[i]);
        //{b iff d contains all the keywords}

        if (!b) return;   //Return with query unchanged

        //{b}
        //{d contains all the keywords}
        //Now update matches as required.

        //CS2:        compute the total occurrence count sum for all keywords
        int sm = 0;
        for(int i = 0; i < keys.length; i=i+1)
            sm = sm + ((Integer) h.get(keys[i])).intValue();
        //{sm = sumAll(d,keys)}

        //CS3:        insert the <d,sum> pair in the ArrayList of matches.
        DocCnt dc = new DocCnt(d,sm);    //construct <d,sum>
        matches.add(dc);                 //add it to matches
        Collections.sort(matches, new SortByCount());  //re-sort matches

    }


    public String[] keys() {
        //EFFECTS: Returns the keywords of this, ie Key

        return keys;   //WARNING: exposes rep
    }


    public int size() {
        //EFFECTS: Returns a count of the documents that match the query

        return matches.size();
    }


    public Doc fetch(int i) throws IndexOutOfBoundsException {
        //EFFECTS: If $0 \le i < \size$ returns the $i$'th matching
        //   document else throws IndexOutOfBoundsException

        if (0 <= i && i < matches.size())
            return ((DocCnt) matches.get(i)).getDoc();
        else
            throw new IndexOutOfBoundsException("Query.fetch");
    }


    public String toString() {
        String s;

        s = matches.size() + " matches:\n";  
        //	for(int i=0; i < matches.size(); i=i+1)
        //	    s = s + (((DocCnt) matches.get(i)).getDoc()).getTitle() + "\n";

        for(int i=0; i < matches.size(); i=i+1) {
            DocCnt dc = (DocCnt) matches.get(i);
            Doc d = dc.getDoc();
            String t = d.getTitle();
            int c = dc.getCnt();
            s = s + "Document \"" + t + "\" contains " + c + " matches\n";
        }

        return s;
    }

}
\end{verbatim}

\subsection{Source Code for the WordTable Class}

\begin{verbatim}

public class WordTable {  
    //OVERVIEW: Keeps track of both interesting and uninteresting words. The uninteresting
    //   words are obtained from a private file.
    //   Records the number of times each interesting word occurs in each document.


    Map map;  //Maps each interesting word w to an ArrayList of all its DocCnt objects.
    //(d,c) in w's ArrayList means that w occurs c times in document d.
    //Maps each uninteresting word to null.

    //REPRESENTATION INVARIANT:
    //   (A w : isInteresting(w) : (d,c) in map.get(w) iff c = sumKey(d,w))



    //CONSTRUCTORS

    public WordTable() throws NotPossibleException, FileNotFoundException {
        //EFFECTS: If the file of uninteresting words cannot be read throws NotPossibleException,
        //   else initializes the Wordtable to contain all the words in
        //   the file as uninteresting words.

        map = new HashMap();

        java.io.File nkfile = new java.io.File("src/uninteresting.txt");  
        //private file containing the uninteresting words
        Scanner input = new Scanner(nkfile);

        //We will map uninteresting words to null instead of to an actual DocCnt object.
        String w;
        while (input.hasNext()) {
            w = input.next();
            map.put(w, null);
        }
    }



    //METHODS

    public boolean isInteresting(String w) {
        //EFFECTS: If w is null or a nonword or an uninteresting word
        //    returns false else returns true
        //HELPS: Engine.queryFirst(w), Engine.queryMore(w)

        if ( (w == null) || (!wordOk(w)) ) return false;
        if ( map.containsKey(w) && map.get(w) == null ) return false;
        return true;  //None of the above, so w is interesting
    }


    public Map addDoc(Doc d) {
        //REQUIRES: d is not null
        //MODIFIES:  this
        //EFFECTS: Adds all of the interesting words of d to this
        //   with a count of their number of occurrences. Also returns
        //   a hashtable (docMap) mapping each interesting word in d to its number of occurrences.
        //HELPS: Engine.addDocs(u)


        //Code sketch:
        //1.    Compute the hashtable docMap and also the set of words in d
        //2.    For each word w in d
        //        construct the DocCnt object dc = (d, docMap(w)) and
        //        insert the pair (w, dc) into this
        //3.    Return docMap



        Map docMap = new HashMap();

        Iterator g = d.words();      //Generator for words in d's body
        Set docSet = new HashSet();  //Set to store all interesting words in d's body,
        //to use later to iterate and insert into this

        String word;    //Used to store an individual word in d's body that is being processed.

        //1.    Compute the hashtable docMap and also the set of words in d (store in docSet)
        while (g.hasNext()) {
            word = (String) g.next();                 //next word of d's body to process
            if (isInteresting(word)) {                //if word is interesting, then

                //update its entry in docMap
                Integer ONE = new Integer(1);
                Integer count = (Integer) docMap.get(word);
                if (count == null) {
                    count = ONE; }
                else {
                    int value = count.intValue();
                    count = new Integer(value + 1);
                }
                docMap.put(word, count);
                docSet.add(word);
            }
        }
        //docMap and docSet have the correct values:
        //   docSet consists of all the interesting words in d, and 
        //   every interesting word in d is mapped by docMap to its //   occurrence count in d,


        //2. For each word w in d, i.e., in docSet
        //        construct the DocCnt object dc = (d, docMap(w)) and
        //        insert the pair (w, dc) into this
        Iterator gs = docSet.iterator();
        while (gs.hasNext()) {             //iterate over words in docSet

            word = (String) gs.next();

            //Construct the DocCnt object dc = (d, docMap(word))
            int c = ((Integer) docMap.get(word)).intValue();
            DocCnt dc = new DocCnt(d, c);

            //Append dc to map(word)
            if (!map.containsKey(word))  {      //word not in map, so map(word) is null
                ArrayList<DocCnt> A = new ArrayList<DocCnt>();  //declare empty arraylist A
                A.add(dc);                      //add dc to A
                map.put(word,A);                //insert (word,A) into map, 
                                                //i.e., into the wordTable
            }
            else {    //word is in map
                ArrayList<DocCnt> A = (ArrayList<DocCnt>) map.get(word);  //get map(word)
                A.add(dc);                    //Append dc to it
                map.put(word,A);              //Put updated ArrayList as new map(word)
            }

        }
        //The value of map is updated correctly

        return docMap;

    }





    public ArrayList lookup(String k) {
        //REQUIRES: k is not null.
        //EFFECTS: Returns an ArrayList of DocCnt objects where the
        //  occurrence count of word k in Doc is Cnt.
        //HELPS: Query.Query(wt, w).

        return (ArrayList) map.get(k);  //WARNING: exposes rep. OK for internal use only.
    }



    public boolean wordOk(String s) {
        //EFFECTS: returns true iff s consists entirely of alphabetic characters.

        for(int i = 0; i < s.length(); i = i+1) {
            char c = s.charAt(i);
            if (!( ('a' <= c && c <= 'z') || ('A' <= c && c <= 'Z') ))  
            //c is not alphabetic so return false immediately
                return false;
        }
        return true;  //All chars in s are alphabetic so return true

    }
}
\end{verbatim}

\subsection{Source Code for the DocCnt Class}

\begin{verbatim}

public class DocCnt {
    //OVERVIEW: Implements DocCnt objects, which are a pair (d,c) consisting of
    //      a document (Doc) d and a count (int) c.
    //      Indicates that some particular word occurs c times in document d.

    Doc doc;
    int count;

    //constructors

    public DocCnt(Doc d, int c) {
        //EFFECTS: Initializes this to (d, c)
        doc = d;
        count = c;
    }

    //methods

    public Doc getDoc() {
        //EFFECTS: returns this.doc, i.e, the document of the DocCnt object.

        return this.doc;
    }


    public int getCnt() {
        //EFFECTS: returns this.count, i.e, the count of the DocCnt object.

        return this.count;
    }

    public String toString() {

        String s;
        s = doc.getTitle() + " doc.getCnt()";
        return s;
    }

}


\end{verbatim}

\subsection{Source Code for the Doc Class}

\begin{verbatim}
public class Doc {

    //OVERVIEW: A document contains a title and a text body. Doc
    //   is immutable and provides an iterator.
    //   A document is constructed from a single string, (which is e.g., the entire
    //   contents of a text file). We make the convention that the first
    //   line of the string (i.e. up to \n) is the title, and the remaining lines are the body,
    //   with whitespace replaced by a single blank.

    String title;    //title of the document
    String body;     //body of the document


    //CONSTRUCTORS

    public Doc(String s) throws NotPossibleException {
        //EFFECTS: if s cannot be processed as a document throws NotPossibleException,
        //   else makes this be the Doc corresponding to s.

        Scanner input = new Scanner(s);

        title = input.nextLine();          //title is first line.

        body = "";             //body is remainder. Replace all whitespace by single blank.
        while (input.hasNext())
            body = body + input.next() + " ";
    }


    //METHODS

    public String getTitle() {
        //EFFECTS: Returns the title of this.
        return(title);
    }


    public String getBody() {
        //EFFECTS: Returns the body of this.
        return(body);
    }



    //ITERATOR

    public Iterator words() {
        //EFFECTS: Returns a generator that will yield all the words in the document
        //as strings in the order they appear in the text.

        return new WordGenerator(this);
    }


    // inner class
    private static class WordGenerator implements Iterator {

        private Doc doc; // the Doc
        Scanner inpt;    //Scanner used to iterate over the body of doc

        WordGenerator(Doc idoc) { //the constructor of the generator
            // REQUIRES: idoc != null

            doc = idoc; //set to the Doc being iterated over,
            //which is passed to the Iterator as "this" in
            //the call to the Iterator.
            //The iterator then passes it on to the generator.

            inpt = new Scanner(doc.body);

        }

        public boolean hasNext( ) {
            //If the inpt scanner still has input, then return true
            return(inpt.hasNext());
        }

        public Object next( ) throws NoSuchElementException {
            //If inpt scanner still has input, then return the next token as a string,
            //otherwise throw exception. Note that we need a wrapper
            //class (String) since the return type of next is Object.

            if ( inpt.hasNext() ) {
                return( inpt.next());
            } else {
                throw new NoSuchElementException("Doc.words");
            }

        }

        public void remove( ) { return; }//WARNING: DOESNT DO ANYTHING, i.e is a stub

    }

}
\end{verbatim}

\subsection{Source Code for the TitleTable Class}

\begin{verbatim}
public class TitleTable {
    //OVERVIEW: Keeps track of documents with their titles.

    Map map; //maps each title (a string) to its matching Doc.
    //We assume that titles are unique (formalize as REQUIRES)

    //constructors

    public TitleTable() {
        //EFFECTS: Initializes this to be an empty table.
        map = new HashMap();
    }

    //methods

    public void addDoc(Doc d) throws DuplicateException {
        //REQUIRES: d is not null.
        //MODIFIES:  this.
        //EFFECTS: If a document with d's title is already in this throws DuplicateException,
        //   else adds d with its title to this.
        //HELPS: Engine.addDocs(u)

        if (map.containsKey(d)) throw new DuplicateException("TitleTable.addDoc");
        else map.put(d.getTitle(), d);
    }


    public Doc lookup(String t) throws NotPossibleException {
        //EFFECTS: If t is null or there is no document with
        //   title t in this throws NotPossibleException
        //   else returns the document with title t.
        //HELPS: Engine.findDoc(t)

        if (t==null || !map.containsKey(t)) throw new NotPossibleException("TitleTable.lookup");
        else return((Doc) map.get(t));
    }
}
\end{verbatim}


\section{Code and Comment Analysis using Claude}

We used the Claude LLM to check if comments and their corresponding code are consistent, \ie if the comment describes the code accurately. We show two queries and the responses that Claude gave.

\subsection{First query}

The first query checks a comment versus code that accurately implements the comment. The response is given verbatim from Claude with some
reformatting for clarity.

\subsubsection{Query to Claude}

\begin{verbatim}
Does the comment at the top match the following code

void addDoc(Doc d, Map h) {
        //REQUIRES: d is not null and h maps strings (the
        //   interesting words in d) to integers (the occurrence
        //   count of the word in d)
        //MODIFIES: this
        //EFFECTS: If each keyword of this is in h,
        //              adds d to the matches of this.
        //HELPS: Engine.addDocs(u), Engine.addDocFromFile(f)
        //Code sketch:
        //CS1:   If the map h for document d contains all the keywords,
        //CS2:        compute the total occurrence count sum for all keywords and
        //CS3:        insert the <d,sum> pair in the ArrayList of matches.
        //CS4:    Otherwise leave matches unchanged.
        //CS5:    In either case, return the resulting query.


        //CS1:   If the map h for document d contains all the keywords,
        boolean b = true;
        for(int i = 0; i < keys.length; i=i+1)
            b = b && h.containsKey(keys[i]);

        if (!b) return;   //Return with query unchanged


        //CS2:        compute the total occurrence count sum for all keywords
        int sm = 0;
        for(int i = 0; i < keys.length; i=i+1)
            sm = sm + ((Integer) h.get(keys[i])).intValue();

        //CS3:        insert the <d,sum> pair in the ArrayList of matches.
        DocCnt dc = new DocCnt(d,sm);    //construct <d,sum>
        matches.add(dc);                 //add it to matches
        Collections.sort(matches, new SortByCount());  //re-sort matches

    }
\end{verbatim}

\subsubsection{Response from Claude}
\begin{verbatim}
The provided code matches the comment at the top. The comment outlines the requirements,
effects, and a code sketch for the addDoc method, which the code adheres to.
Specifically:

1. The comment states the method requires d to be non-null and h to be a map from 
strings (keywords) to integers (occurrence counts).

2. The comment mentions the method modifies the current object (this).

3. The effect outlined is that if all keywords of the current object are present in h, 
then d is added to the matches of the current object.

4. The code sketch outlines the steps:
CS1: Check if h contains all keywords
CS2: Compute the total occurrence count sum for all keywords
CS3: Insert the <d, sum> pair into the matches list
CS4: Otherwise, leave matches unchanged
CS5: Return the resulting query

The code follows this sketch:

CS1 is implemented by iterating over keys and checking if each key is present in h.
If any key is missing, b becomes false, and the method returns without modifying matches (CS4).
CS2 is implemented by iterating over keys and summing the occurrence counts from h.
CS3 is implemented by creating a DocCnt object with d and the sum, then adding it to 
matches and sorting matches.

So the code correctly implements the behavior described in the comment.
\end{verbatim}

\subsection{Second query}

The second query checks a comment versus code that does not accurately implement the comment.
The response is given verbatim from Claude with some
reformatting for clarity.
We have included, along with in-line code comments in natural language, Hoare-logic style assertions. The presence of these appears to help Claude produce the correct response.

We will in future work investigate principles for combining natural language and formal assertions so as to maximize the accuracy of the responses of the LLM.

\subsubsection{Query to Claude}

\begin{verbatim}
Does the comment at the top match the following code

void addDoc(Doc d, Map h) {
        //REQUIRES: d is not null and h maps strings (the
        //   interesting words in d) to integers (the occurrence
        //   count of the word in d)
        //MODIFIES: this
        //EFFECTS: If each keyword of this is in h,
        //              adds d to the matches of this.
        //HELPS: Engine.addDocs(u), Engine.addDocFromFile(f)
        //Code sketch:
        //CS1:   If the map h for document d contains all the keywords,
        //CS2:        compute the total occurrence count sum for all keywords and
        //CS3:        insert the <d,sum> pair in the ArrayList of matches.
        //CS4:    Otherwise leave matches unchanged.
        //CS5:    In either case, return the resulting query.


        //CS1:   If the map h for document d contains all the keywords,
        boolean b = true;
        for(int i = 0; i < keys.length; i=i+1)
            b = b && h.containsKey(keys[i]);
        //{b iff d contains all the keywords}

        if (b) return;   //Return with query unchanged

        //{b}
        //{d contains all the keywords}
        //Now update matches as required.

        //CS2:        compute the total occurrence count sum for all keywords
        int sm = 0;
        for(int i = 0; i < keys.length; i=i+1)
            sm = sm + ((Integer) h.get(keys[i])).intValue();
        //{sm = sumAll(d,keys)}

        //CS3:        insert the <d,sum> pair in the ArrayList of matches.
        DocCnt dc = new DocCnt(d,sm);    //construct <d,sum>
        matches.add(dc);                 //add it to matches
        Collections.sort(matches, new SortByCount());  //re-sort matches

    }
\end{verbatim}

\subsubsection{Response from Claude}

\begin{verbatim}
No, the code does not match the comments at the top. There is a mistake in the 
implementation of the step described in CS4:

**CS4: Otherwise leave matches unchanged.**

In the code, the condition is reversed. Instead of leaving the `matches` unchanged if 
the document `d` does not contain all the keywords, the code returns immediately when
`d` contains all the keywords:

```java
if (b) return; //Return with query unchanged
```

This means that if `d` contains all the keywords, the method exits without updating 
the `matches` list. However, according to the comment, the `matches` list should be updated
with the `<d, sum>` pair when `d` contains all the keywords.

The correct implementation of CS4 should be:

```java
if (!b) return; //Return with query unchanged
```

This way, the method returns without modifying `matches` when `d` does not contain 
all the keywords, as described in the comment.

The rest of the code seems to match the comments correctly:

- CS1 checks if `d` contains all the keywords.
- CS2 computes the sum of occurrence counts for all keywords.
- CS3 inserts the `<d, sum>` pair into the `matches` list and sorts it.

However, due to the mistake in the implementation of CS4, the code does not fully
match the comments at the top.
\end{verbatim}




\clearpage
\bibliographystyle{plain}
\bibliography{refs}

\end{document}